# Nanoscale Substrate Roughness Hinders Domain Formation in Supported Lipid Bilayers

*James A. Goodchild, Danielle L. Walsh, Simon D. Connell\**

School of Physics and Astronomy, University of Leeds, Leeds, United Kingdom

## Abstract

Supported Lipid Bilayers (SLBs) are model membranes formed at solid substrate surfaces. This architecture renders the membrane experimentally accessible to surface sensitive techniques used to study their properties, including Atomic Force Microscopy (AFM), optical fluorescence microscopy, Quartz Crystal Microbalance (QCM) and X-Ray/Neutron Reflectometry, and allows integration with technology for potential biotechnological applications such as drug screening devices. The experimental technique often dictates substrate choice or treatment, and it is anecdotally recognised that certain substrates are suitable for the particular experiment, but the exact influence of the substrate has not been comprehensively investigated. Here, we study the behavior of a simple model bilayer, phase separating on a variety of commonly used substrates, including glass, mica, silicon and quartz, with drastically different results. The distinct micron scale domains observed on mica, identical to those seen in free-floating Giant Unilamellar Vesicles (GUVs), are reduced to nanometer scale domains on glass and quartz. The mechanism for the arrest of domain formation is investigated, and the most likely candidate is nanoscale surface roughness, acting as a drag on the hydrodynamic motion of small domains during phase separation. Evidence was found that the physico-chemical properties of the surface have a mediating effect, most likely due to changes in the lubricating interstitial water layer between surface and bilayer.

# Introduction

Model lipid membranes have been studied extensively to investigate the fundamental structure and physics of the cell membrane,[1–6] understand protein and drug interactions with the membrane,[7] and to develop biotechnological applications such as drug delivery systems.[8] The ability to form Supported Lipid Bilayers (SLBs) on solid substrates renders them experimentally accessible to surface sensitive techniques, such as Atomic Force Microscopy (AFM),[9] Quartz Crystal Microbalance with Dissipation (QCM-D),[10] and fluorescence techniques such as Fluorescence Correlation Spectroscopy (FCS)[11,12] and Fluorescence Recovery After Photobleaching (FRAP),[13,14] yielding information about lipid diffusion, ordering, structure and phase behaviour.[9,12,15,16]

Many substrates can be used to support lipid bilayers including mica, glass and silica, the choice usually driven by the signal being measured. AFM predominantly uses mica, a mineral that is easily cleaved to be atomically flat, enabling clear discrimination of small bilayer height steps.[2,3] Fluorescence microscopy techniques are best utilised using glass due to its optical transparency.[14,17] Issues arise when comparing results between different surface sensitive techniques due to the different substrates used. It is therefore imperative that the effect of substrate is understood and taken into account when interpreting and comparing the data arising from different experiments, between surface and free-floating bilayers, between different surface-sensitive instruments, and between research groups.

Strategies to decouple the bilayer from the substrate include double bilayers,[13,18,19] tethering of a free floating bilayer to a surface,[20] and supporting the bilayer on a hydrated polymer cushion.[21] These methods can be effective but they increase the complexity of the sample preparation, and are only suitable for certain techniques.

Whilst it is recognised that substrates can affect bilayer properties compared to free-floating systems, the extent of the effect, the influence of experimental conditions, and even the interaction mechanism are not understood. Although there is a thin interstitial water layer between the bilayer and the substrate which allows the bilayer to remain fluid, the lipid diffusion drops for both mica and glass SLBs compared to free standing Giant Unilamellar Vesicles (GUVs) and Black Lipid Membranes (BLMs).[22,23] Phase separation can also change significantly in SLBs. This is important because the separation of immiscible lipid types into distinct phases with separate ordering, diffusion and lateral density has been linked with the potential existence of lipid rafts (or nano-domains) in the plasma membrane, which are widely studied.[1] Phase domains on solid supports can vary in shape and size compared to GUVs, and in particular coalescence of domains appears to be hindered due to a surface interaction.[16,24]

Phase behaviour has been well characterised on mica, to give information on domains such as size, height and dynamics.[2,4,25] However, reports of phase separation on glass are scarcer. Domains have been observed on glass following Langmuir-Blodgett Deposition,[16,26,27] where the domains are already present at the liquid-air interface before deposition, and also in phase separated GUVs ruptured onto glass.[16,17,28] In the literature we have only found a few studies showing domains forming on glass via vesicle fusion, where the domains would have to nucleate and grow from a single homogenous phase on the substrate.[11,14,29] This is remarkable considering the ubiquitous use of glass in optical microscopy,[15,30,31] and the hundreds if not thousands of papers showing phase separation in free-floating GUVs,[5,17,26] and in SLBs on mica.[2,3,24,25] There are also relatively few reports showing domain formation on Si/SiO$_2$ substrates via vesicle fusion, but micron size domains have been observed on this substrate.[32,33]

In this study we find that DPPC/DOPC (60:40) bilayers form irregular nanoscale domains on glass and other substrates that are rough at the nanoscale, in contrast to micron scale fractal domains on smoother mica and silicon. These nanoscale domains are beyond the resolution of diffraction limited microscopy and are resolved by AFM. We find no difference in molecular diffusion on the different substrates, and together with multiple other strands of evidence we propose a significant hindering of hydrodynamic lipid flow resulting in restricted domain growth. Increased roughness is the likely cause of this restriction in hydrodynamic lipid flow but mediated by chemical differences via the interstitial water gap.

# Experimental Section

## Preparation of Lipid Vesicles

DOPC (1,2-dioleoyl-*sn*-glycero-3-phosphocholine), DPPC (1,2-dipalmitoyl-*sn*-glycero-3-phosphocholine) and 16:0 NBD PE (1,2-dipalmitoyl-sn-glycero-3-phosphoethanolamine-N-(7-nitro-2-1,3-benzoxadiazol-4-yl, ammonium salt) were purchased from Avanti Polar Lipids (Alabaster, AL). Texas Red DHPE (Texas Red 1,2-Dihexadecanoyl-sn-Glycero-3-Phosphoethanolamine, Triethylammonium Salt) was purchased from Thermo Fisher Scientific UK. DOPC, DPPC, Texas Red DHPE and 16:0 NBD PE were dissolved into individual 5mM CHCl$_3$ stock solutions and then mixed together in the desired composition, dried under nitrogen for 20 min and then kept under vacuum overnight. The dry film was hydrated in ultrapure water (Milli-Q), vortexed for 30 min, heated in an oven at 50 °C for 30 min and then tip sonicated for 30 min at 4 °C. The resulting Small Unilamellar Vesicle (SUV) sample was then centrifuged at 3000rpm for 3 min, to separate SUVs from any metal sediment from the sonicator tip.

## Substrate Preparation

Round glass coverslips (Thermo Scientific, Menzel-Glaser, 'Glass 1') were prepared by bath sonicating in Decon detergent for 15 min, followed by 10 min piranha treatment (1:3 $H_2O_2$:$H_2SO_4$), followed by 20 min exposure to UV ozone (UVOCS Inc. UV Ozone Cleaning System). These coverslips were used as they were for the fluorescence flow cell and glued to a magnetic stub using 2-part epoxy to be used on the AFM stage. 'Glass 2' coverslips (VWR borosilicate), Quartz slides (Alfa Aesar) and Silicon Wafers (polished, <100>, B dopant, Flats: SEMI std., purchased from Inseto, UK) were prepared using the same procedure. Throughout this paper glass refers to Glass 1 (Menzel Glazer) and Glass 2 (VWR) is always referred to as Glass 2. Thermally oxidised silicon (polished, 100 nm thick wet thermal oxide, <100>, N/Phos dopant, Flats: 2 SEMI, purchased from Si-Mat, Kaufering, Germany) was either prepared by cleaning with acetone and IPA (no plasma) or by additionally oxidising using a Diener Electronic Zepto Oxygen Plasma Laboratory unit for 2 min at 0.6-0.8 mBar (100W, 40kHz).

Mica (Agar Scientific) substrates were cleaved using adhesive tape immediately prior to use. The mica was cut to size to fit into the fluorescence fluid cell, and for AFM was glued to a magnetic stub with 2-part epoxy.

To etch mica, stubs were cleaved and placed in PTFE beakers with 40% Hydrofluoric Acid (HF). Pit depth and density were controlled by systematically varying the etch time, from 1 min to 4 hr. An ideal time of 30 min produced pits of 1.0 nm depth defined by the cleavage planes of mica, with only a small number of 2nd layer pits initiating within the larger pits. The beaker and mica stubs were then poured into a large amount of Sodium Bicarbonate (90g in 1L), and then thoroughly rinsed with de-ionised water. This method was originally developed as a calibration tool for AFM.[34] A colleague trained in the safe handling of HF performed this procedure.

## Supported Bilayer Formation

For fluorescence measurements, glass or mica substrates were assembled into a home-built flow cell consisting of a sealed incubation chamber and liquid inlet and outlet. 1mL of 1mg/mL lipid vesicles were syringed into the cell. The vesicles were incubated on the surface for 30 min (room temperature for DOPC, 50 °C for DPPC/DOPC and pure DPPC - to ensure all lipid is in the fluid state, freely mixed in one phase that will adsorb to the surface with equal probability with no sorting effect). 1 mL 20 mM $MgCl_2$ at the same temperature was added for a further 30 min. The sample was then allowed to cool to room temperature before the surface was washed with Milli-Q to remove $Mg^{2+}$ and any unfused vesicles. A pump was connected to flow room temperature Milli-Q water through at approximately 1 mL/min for 30 min.

For AFM measurements, 100 µL of SUV solution was deposited onto a freshly cleaved mica disk or onto freshly cleaned glass/quartz/silicon, and incubated in a sealed humidity chamber for 1h at 50 °C. Halfway through incubation 100 µL 20 mM $MgCl_2$ was added. After incubation, the bilayer was cooled to room temperature and rinsed to remove $Mg^{2+}$ and any unfused vesicles by exchanging with Milli-Q at the same temperature as the incubated bilayer, washing across the surface 10 times in 100 µL bursts.

The role of $Mg^{2+}$ is to aid vesicle fusion and produce the high-quality defect free bilayers preferable for AFM. It works by enhancing lipid-lipid interactions, and appears to have no effect on domain size, either through a condensing effect, or by co-ordination with surface moieties across the interstitial water gap. Protocols vary according to lipid chemistry and state. An example of this lipid mixture prepared without $Mg^{2+}$ can be seen in Figure 6.

The temperature of the bilayers was measured using a K-type thermocouple positioned close to the substrate. Two separate cooling rates from 50 °C to room temperature (21 °C) were achieved by turning off the oven or by taking the samples out of the oven, 0.080±0.008 °C/Min (N=4) and 0.25±0.02 °C/min (N=3) respectively. The cooling rates were determined by taking the gradient between 33-29 °C , this is the transition temperature ($T_m$) range of DPPC/DOPC (60:40), determined using published DPPC/DOPC temperature Phase Diagrams.[35,36] The cooling rate for pure DPPC samples was calculated between 45-35 °C to match the cooling rate at $T_m$ of pure DPPC (40.73±0.03 °C, see DSC SI).

## Fluorescence and FRAP

Fluorescence Microscopy was performed using a Nikon Eclipse E600 microscope with an Andor Technology Zyla cCMOS camera. The microscope was equipped with a Mercury Lamp, filter cubes suitable for Texas Red (Ex. 540-580, Em. 600-660) and NBD (Ex. 465-495, Em. 515-555), and x40 air and x100 oil objectives.

For Fluorescence Recovery after Photobleaching (FRAP) measurements, an aperture was used to bleach a 30 µm diameter spot with white light for 30 s. After photobleaching, images were taken at 3 s intervals for 3 min. Analysis was performed using a custom macro written for ImageJ, which compares the fluorescence intensity recovery to a reference area of non-bleached bilayer. The exponential recovery was fitted to obtain a recovery half-life ($t_{1/2}$), which can then be converted to a diffusion coefficient (D).

$$D = \gamma_D \left( \frac{r^2}{4t_{\frac{1}{2}}} \right)$$

Where r is the radius of the bleach spot and $\gamma_D$ is a constant (0.88) related to the circular bleach shape.

The diffusion coefficient values are averages of several repeat runs (Glass N=12, Mica N=6), where for each repeat run the value is an average of at least 5 different areas from the substrate. T-tests were performed using SPSS (IBM) Software.

For transition temperature determination, DPPC bilayers were formed in the flow cell, as described earlier. After the wash at room temperature, the bilayer was heated up to 60 °C and then FRAP images obtained as it cooled. Diffusion Coefficient (D) vs. Temperature (T) plots were fitted to a Boltzmann Sigmoidal curve.

$$D = A_2 + \frac{A_1 - A_2}{1 + e^{\frac{T - T_o}{dT}}}$$

Where A1 and A2 are the y values of the flat fit above and below sigmoid curve and $T_o$ is the turning point/midpoint of the curve, which is taken as value of $T_m$. $T_m$ values are averages of repeat runs (Glass N=5, Mica N=4).

FRAP was performed using equipment available and within time-constraints of taking measurements during cooling. Ideally, bleach time should be short compared to $t_{1/2}$, but the microscope lamp used required 30s to fully bleach the bilayer, so fluorescence recovery had commenced during bleaching. In particular, D values for the solid phase would not be accurate due to the short acquisition time. Although the D values stated will be incorrect to an absolute standard, they are adequate to compare the diffusion on the different substrates using the same membrane under identical conditions, *and* allow the surface thermal transition temperature to be accurately determined. In addition, our diffusion values agree with literature values from different techniques, which vary between 0.5-5.0 µm$^2$/s for fluid lipid systems.[12,22,31,37]

## Atomic Force Microscopy (AFM)

AFM Images were acquired using a Bruker Dimension ICON. Bruker ScanAsyst-Fluid (0.7 N/m, 150 kHz) were used for imaging bilayers using Peak Force Tapping in liquid. It is well established that in tapping mode (amplitude-modulation AFM), the height signal can be influenced in the nanometer range by heterogeneous sample compressibility and differential damping of the oscillating probe.[38,39] Peak force tapping is preferred because the feedback mechanism provides direct control of the applied force, which is generally at the minimum to achieve stable imaging, around 150-300 pN. Bruker FastScan A probes (18 N/m, 1.4 MHz) were used for imaging all of the clean substrates using high speed Tapping Mode in air, and for some of the bilayers (Silicon, thermally oxidized silicon and quartz), FastScan D probes (0.25 N/m, 110 kHz) were used in liquid. Images were acquired at a minimum 768x768 pixel resolution.

Z-noise is 30 pm, and this response can be seen in Figure 5 in the measurement of roughness on an atomically smooth cleavage plane of mica. Although ultimate lateral resolution can be 1 nm, on lipid bilayers the smallest lipid domains we can visualize are around 5 nm in diameter, equating to approximately 40 lipid molecules. However, in another simulation study the size of the transition between $L_o$-$L_d$ phases is also about 5 nm wide,[40] and together with tip convolution the possibility arises that AFM is detecting the smallest clusters that could conceivably be defined as a phase.

## Image Analysis

Fluorescence microscopy images were analysed and processed using the FIJI distribution of ImageJ (NIH). AFM images were analysed using Nanoscope Analysis v1.9. AFM images were flattened using the appropriate lowest order of levelling for each image. $R_a$ roughness was measured from multiple images located across the sample, all at the same 5 μm scan size as this can affect the $R_a$ measured. A more informative quantitative measure of roughness at different length scales was measured using the Power Spectral Density (PSD) function (Figure 5C), plotted as the power in the surface corrugation vs log of the lateral wavelength. Domains sizes were estimated by fitting an ellipse to the domain using ImageJ automated particle analysis or Nanoscope Analysis, and then taking the average of the long and short radii of the ellipse (Figure S1).

For nanoscale domains that were less defined as separated domains, automated particle analysis gave unreliable results, so image correlation was used to extract a characteristic length scale. The processed AFM images were converted to binary using the threshold tool in ImageJ, then a Radially Averaged Autocorrelation Function Macro (Michael Schmid, 27/9/2011 update) was applied to produce an autocorrelation plot, giving the length scale between black and white pixels i.e. the two different phases. This plot was fitted to an exponential decay using Origin Pro 9.1 to give a characteristic correlation length.

$$f(r) = Ae^{\frac{-r}{\xi}}$$

Where f(r) is the autocorrelation, r is distance and $\xi$ is correlation length. Figure S2 shows the process of calculating correlation length from an AFM image. Extra detail on correlation length and how it compares to domain fitting can be found in the SI.

## Contact Angle

Static Contact Angle measurements were taken using a First Ten Angstroms FTA 4000 CAG. An approximately 0.2µL droplet of Milli-Q water was pipetted onto the surface and an image captured. The contact angle of the water with the mica and glass substrates was measured using fitting algorithms in the FTA 400 CAG software, to give the contact angle made between the surface and droplet. For mica N=18 (three repeats on six mica stubs) and for glass N=9 (three repeats on three glass cover slips). The nominal instrumental uncertainty is $\pm$ 2°,[41] but replicate measurements were more reproducible.

# Results

## Phase Structure is Different on Mica and Glass

To investigate the effect of substrate on phase separation, DPPC/DOPC (60:40) bilayers were formed on both freshly cleaved mica, and on Piranha and UV Ozone cleaned glass, under identical incubation conditions. They were imaged using AFM and optical fluorescence microscopy, with the bilayers on glass containing a small proportion of dye labelled lipid (Texas Red or NBD, as indicated) and the bilayers produced for AFM imaging being label free. Domains on mica are large (several micrometres across) with regular flower or finger-like morphologies whose formation has been associated with molecular tilt in the crystallised domains. AFM can distinguish the nanoscale variation in bilayer height between the tightly packed, taller gel phase and the shorter fluid phase (Figure 1A,D,H). This characteristic flower-like morphology is replicated in fluorescence microscopy, where TR-DHPE is excluded from the tightly packed gel domains (dark areas) (Figure 1C+G). This morphology matches closely to DPPC/DOPC domains observed on mica in the literature.[25] However, when the same lipid mixture was incubated under identical conditions on the glass substrate, fluorescence images showed no clear phase separation (Figure 1E+I) because the domains, as observed by AFM, are approximately two orders of magnitude smaller, and hence below the diffraction limit (Figure 1B,F,J). Visually they also show quite a broad distribution of sizes in contrast to the regular size on mica and show rough irregular domain boundaries.

Identification of nanoscale phase separated lipid domains is straightforward as the step heights between the liquid and solid phases are consistent, with the height histogram resolving into a bimodal distribution, whereas the clean surface has a continuously varying height and length scale. Additional mechanical signal channels in the AFM support the assignment (quantitative elastic modulus in Peak Force mode, or a qualitative phase contrast image in tapping mode) and these clearly discriminate the material property of ordered domains from the surrounding fluid phase (Figure S10).

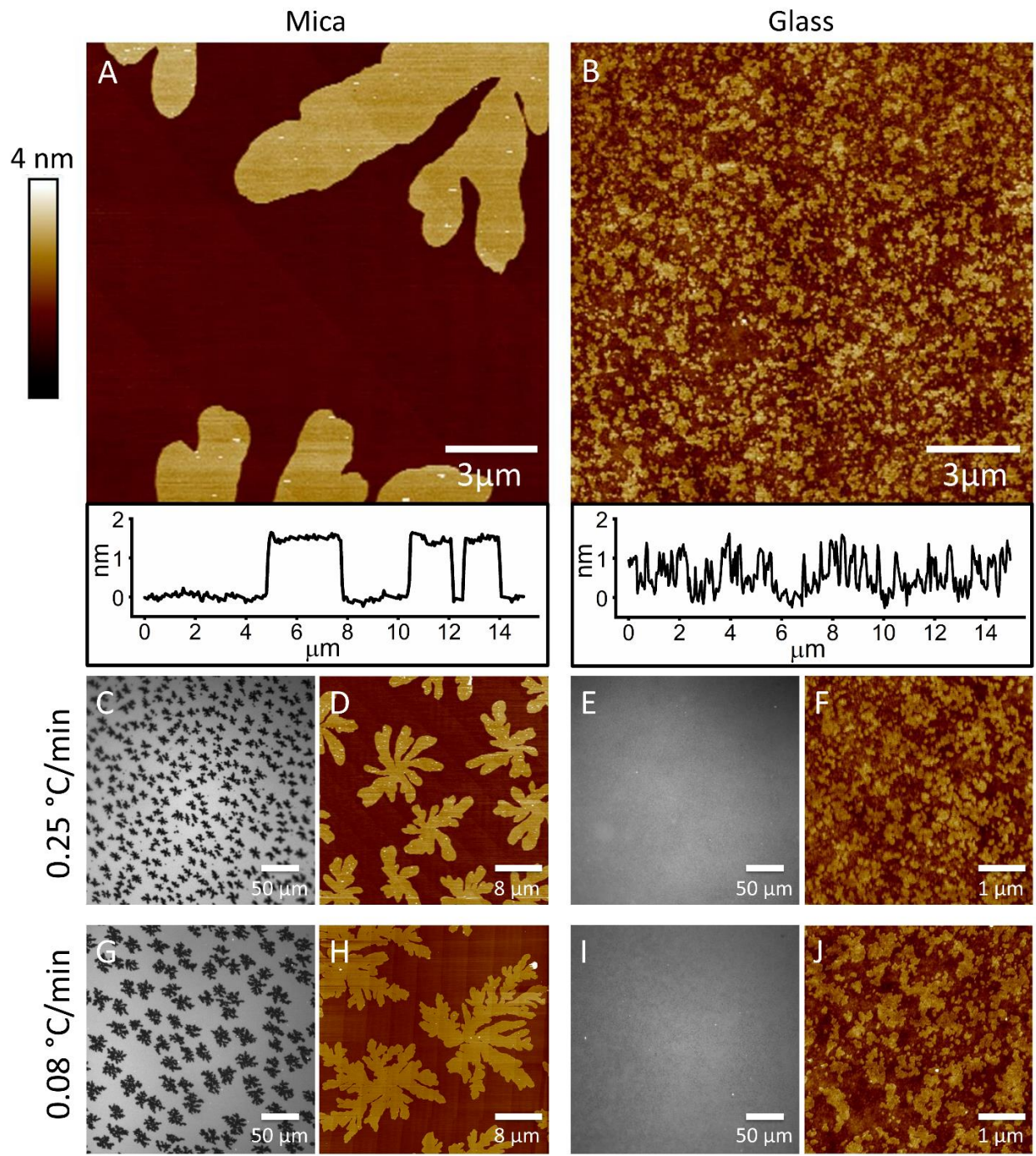

**Figure 1. DPPC/DOPC (60:40) SLBs.** AFM images on mica (A) and glass (B) at the same x,y and z scale, highlighting the discrepancy in size and morphology of domains. Line scans

across the full width of images A and B are included to aid interpretation. AFM images of DPPC/DOPC (60:40) SLBs (D,F,H,J) and fluorescence microscopy images of DPPC/DOPC (60:40) + 0.5%TR SLBs (C,E,G,I). The AFM images on glass are representative examples from a heterogeneous surface, with further examples given in Figure S5. The Z range of all the AFM images is 4 nm, for reference the step height of domains in image B = 1.1 nm. The cooling rates from incubation temperature to room temperature are shown on the left-hand side and apply to the whole row, A and B are also 0.25±0.02 °C/min.

To quantitatively compare domain sizes on mica between the AFM and fluorescence images, and between mica and glass, the domains were analyzed using automated particle analysis and fitting to ellipses, then taking the average of the long and short axes (Figure S1). Domains on mica measured by AFM and fluorescence were very similar, as expected (Table 1).

Domains imaged on glass, as well as being much smaller, also had a partially interconnected morphology, resulting in difficulties with the automated particle analysis. Instead, a radially averaged correlation length was calculated from binary images, providing a length scale measurement of domains (Table 1 and Correlation Length Section in SI). This analysis was also carried out on images of larger domains to verify the output of the two methods. The average AFM image correlation length of domains on glass was 74±5 nm, but almost 2 orders of magnitude larger for domains on mica at 2.3±0.4 μm (Table 1). Correlation length values tend to be smaller than the equivalent particle analysis values due the finger-like fractal appearance of the larger domains (see Correlation Length SI), but nonetheless give comparable readings.

| Substrate | Domain Radius (AFM) / μm | Domain Radius (Optical) / μm | Correlation Length (AFM) / μm | Correlation Length (Optical) / μm | Substrate Roughness, Ra / nm |
|---|---|---|---|---|---|
| Mica | 5.3 ±0.2 | 4.57 ±0.04 | 2.3 ±0.4 | 3.2 ±0.2 | 0.03 ±0.03 |
| Mica (slow cool) | 8.2 ±0.8 | 8.9 ±0.1 | 3.3 ± 0.1 | 7.3 ±0.1 | " |
| Glass | Analysis fails | < resolution | 0.074 ±0.005 | < resolution | 0.15 ±0.03 |
| Glass (slow cool) | Analysis fails | < resolution | 0.065 ±0.007 | < resolution | " |
| Glass 2 | Analysis fails | - | 0.014 ±0.005 | - | 0.17 ±0.03 |
| Quartz | Analysis fails | - | 0.015 ±0.005 | - | 0.41 ± 0.03 |
| Silicon | 0.45 ±0.04 | - | 0.23 ±0.01 | - | 0.05 ±0.03 |
| Therm Ox Si, no plasma | Analysis fails | - | 0.030 ±0.005 | - | 0.22 ±0.03 |
| Therm Ox Si + plasma | 0.89 ±0.06 | - | 0.55 ±0.04 | - | 0.23 ±0.03 |

**Table 1. Domain Sizes and Correlation Lengths for Mica and Glass bilayers at different cooling rates.**

**Cooling rates are all ambient (0.25±0.02 °C/min) except where specified as slow cool (0.080±0.008 °C/min). Domain Fitting was not possible on nanoscale domains due to irregular and interconnected morphology. Optical measurements were not carried out in lower half of table due to nanoscale domains or opaque substrate.**

A close inspection of the fluorescence images of bilayers on glass showed a fine speckled structure (Figure 1E+I), seen more clearly when magnified (Figure S3C). To verify if this was a result of the nanoscale domains being viewed through diffraction limited optics, AFM images of domains on glass were converted to greyscale with the same fluorescence intensity values as the larger scale domains, then blurred using a 500 nm Gaussian filter (500 nm being the optical resolution at the illumination wavelength and lens numerical aperture used). The observable pattern in the AFM images now looks very similar to the speckle pattern that is observed in fluorescence images on glass (Figure S3), indicating that it originates in the nanoscale domains at and below the diffraction limit.

## Phase Structure on Other Substrates

A variety of silicate substrates have been used to support lipid bilayers in other studies, so to gain a more comprehensive understanding bilayers were formed under identical incubation conditions on a second type of commonly used borosilicate glass 'Glass 2', quartz glass, silicon, and thermally oxidised silicon (Figure 2 and Table 1). Silicon, after UV ozone or $O_2$ plasma treatment, and thermally oxidized silicon have silica ($SiO_2$) layers at their surface, like glass and quartz. The domains on Glass 2, Quartz and thermally oxidised silicon were again very small, measured using correlation length at Glass 2 = 14±1 nm, Quartz = 15±1 nm and thermally oxidised silicon = 30±1 nm. By contrast, bilayers on plasma treated silicon and plasma treated thermally oxidized silicon resulted in a domain structure more similar to that on mica at 0.2-0.6 μm. Micron scale domains have been observed on $SiO_2$ by vesicle fusion previously.[32,33] In order to explain the origin of this large difference in behavior the remainder of this paper will focus on comparing a single type of glass with mica in more detail.

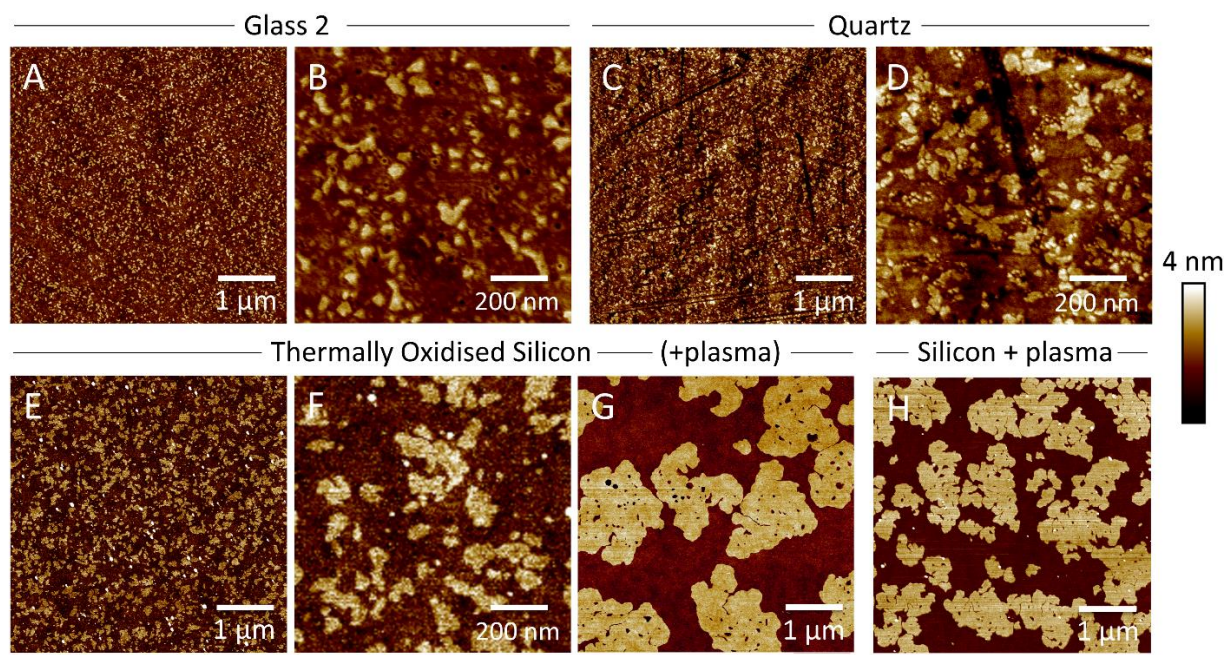

**Figure 2. AFM images of DPPC/DOPC (60:40) SLBs formed on Glass 2, Quartz, thermally oxidised silicon and silicon. Both Glass 2, Quartz and thermally oxidised silicon show nanoscale domains similar to the original glass. It was possible to form micrometer scale domains on thermally oxidised silicon and silicon when the surface was treated with UV ozone or oxygen plasma.**

## Single Lipid Crytallisation is also different on Mica and Glass

Observation of a single component lipid membrane also showed restricted growth of the crystallising domains on glass compared to mica. In DPPC + TR-DHPE and DPPC + 16:0 NBD PC bilayers on mica, the dye-lipids preferentially segregated into the remainder liquid phase as the bilayers cooled from the melt to a solid phase (Figure S4). This embeds a signature of the crystallisation process in the final solid membrane with the domains showing a similar shape as in the mixed DPPC/DOPC system (Figure 3). The Texas Red is a sterically bulky label and is excluded vigorously from the solidifying domains, despite having an identical lipid tail. With the entire membrane solidified, the TR-lipids are concentrated into a narrow boundary around the nucleated domains. In contrast, the NBD label is considerably smaller and therefore excluded from the crystallising solid domains to a lesser degree. This leaves the original dark points of nucleation surrounded by an increasing gradient of dyed-lipid, producing a striking image of the entire process of nucleation and growth.

When DPPC +16:0 NBD PC and DPPC +TR-DHPE were formed on glass, no structure was observed optically. As with the phase separated systems, the glass substrate is hindering the growth of domains.

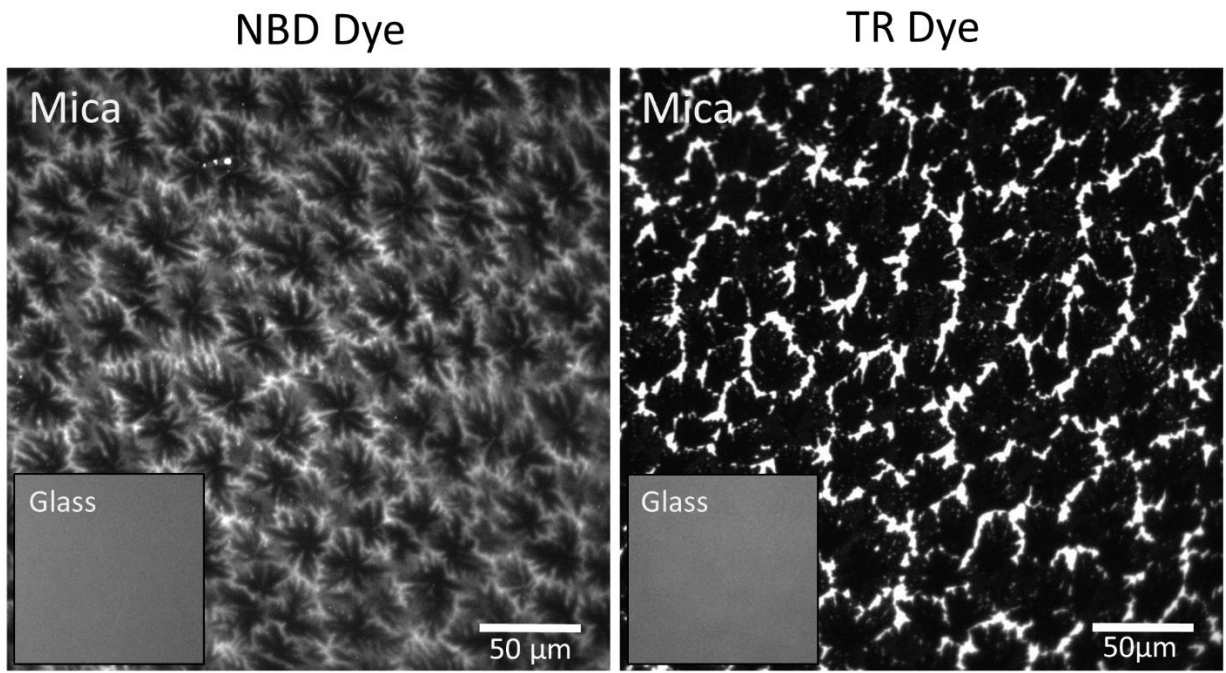

**Figure 3. Room Temperature Images of DPPC with 0.5mol% NBD or 0.5mol% TR.** On mica a signature of the nucleation process remains, the dye molecules are excluded and pushed into the remaining liquid phase. The bulky TR dye is more excluded than the smaller NBD dye. Cropped insets show the same mixtures on glass at the same magnification. On glass no exclusion is observed, likely because it is below the diffraction limit.

## Hydrophobic Forces

To test whether differences in substrate hydrophilicity or the presence of surface contamination could be affecting domain formation, water contact angles were measured. Bilayers will only form if the contact angle is < 30°, and on glass this can be achieved with piranha cleaning. To bring the contact angle of glass close to that of mica = 3±2° (after cleavage), further UV ozone or $O_2$ plasma treatment is required, and a contact angle = 5±2° was achieved (see Table S1 for details). The glass wets well with no apparent pinning points, and bilayers form readily. Recent work on nanobubbles found their formation to be sensitive to surface contamination, the cleaner the surface the more the nanobubbles formed.[42] The sequence of our surface treatment protocol was beyond the extreme end of that used in the nanobubble study, giving confidence that the substrates are as clean as possible, and highly hydrophilic, so that strong interactions with hydrophobic regions of the surface are not causing domain growth the be restricted. This protocol was applied to all substrates studied (except mica).

## Cooling Rates and Domain Size

In mixed lipid systems the influence of lipid diffusion has been shown by changing the cooling rate. By decreasing the cooling rate through the miscibility transition temperature ($T_m$) the domains formed on mica grow larger by allowing more time for lipids to diffuse towards and attach to an expanding nucleating domain.[3,43] In a DPPC/DOPC system, when the cooling rate from incubation temperature (50 °C) down to room temperature was slowed, the size of the gel domains on mica increased both in AFM and fluorescence experiments, confirming this mechanism (Figure 1 C to G and D to H and Table 1). Conversely, the characteristic length scale of the domains on glass does not change as the cooling rate is decreased (Figure 1 F to J, Table 1, Figure S5). Domain size is diffusion limited on mica, but another mechanism is controlling domain size on glass and hindering the growth of domains to a limiting length scale.

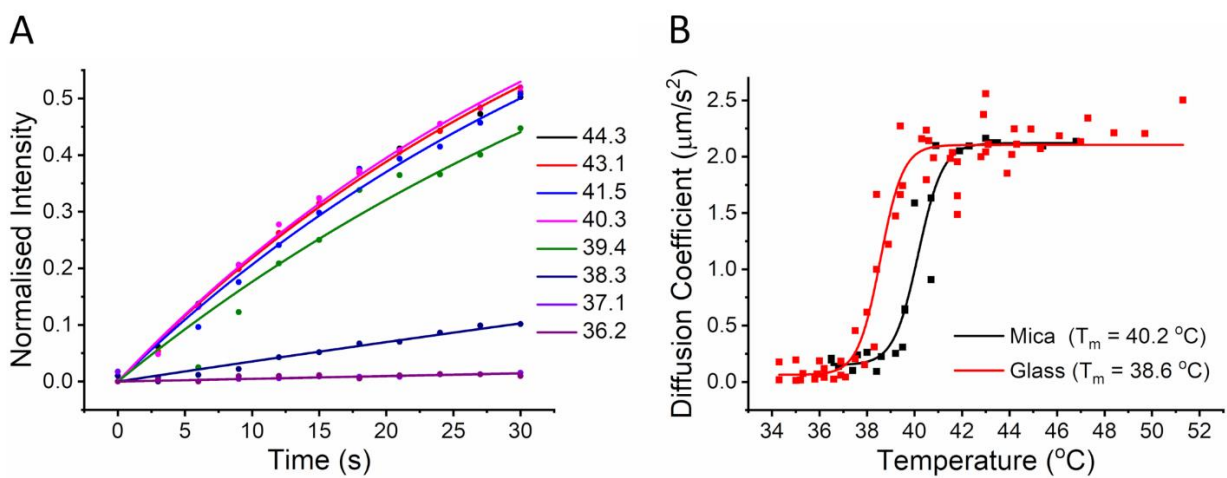

**Figure 4. Transition Temperature Determination. A) An example run of FRAP fluorescence recovery over time curves for DPPC+ 0.5mol% NBD SLB on glass as the bilayer cools, with exponential recovery fit B) Calculated Diffusion coefficients (D) at each temperature for mica and glass, plotted against temperature. For Mica 4 repeat runs are plotted, for glass 5 repeat runs. Data fitted to a Boltzmann sigmoid, $T_m$ value taken as midpoint of sigmoid. $T_m$ values are averages of all repeats.**

## Molecular Diffusion

Whilst the relationship of domain sizes with cooling rate indicated molecular diffusion is not a factor, it is also a secondary measure and a direct measurement is required. Whilst FCS experiments have found that there is no significant difference between DOPC diffusion on glass and mica,[12] for SLBs formed using the same vesicle fusion bilayer deposition method as in this paper, there are many varied results in the literature.[22,31,37] To verify the finding of an identical diffusion constant on mica and glass in our experimental set-up we used FRAP on DOPC + 0.5mol% TR-DHPE bilayers at RT. There was no significant difference between diffusion on mica (0.96±0.04 μm²/s) and glass (1.02±0.04 μm²/s) [Independent samples t-test, mica (M=0.96, SD=0.10), glass (M=1.02, SD=0.10), conditions t(13.47)=1.040,p=0.31]. This was also true for the higher $T_m$ DPPC in the fluid phase above $T_m$ (Figure 4). A DPPC SLB was cooled from 50 °C down to room temperature performing FRAP at regular intervals (Figure S7). The diffusion for fluid phase DPPC + 0.5mol% 16:0 NBD PE above $T_m$ on mica was 2.1±0.1 μm²/s and on glass was 2.3±0.3 μm²/s.

Some of the variation in diffusion coefficient in the literature could indicate that the probe molecule being measured is not purely reporting on the state of the adjacent bilayer, but is also influenced by its own interaction with the substrate or interstitial water later, and it does not precisely report the absolute lipid diffusion rate. Experiments to measure the difference between TR and NBD were not conclusive but it is possible that the larger fused heterocyclic moiety of the TR dye labelled lipid experiences drag in this way. Most importantly for this study, the low concentration (0.5%) of dye label does not appear to have affected the formation of domains, with the AFM showing very similar sized domains on an unlabeled lipid system.

By fitting the change in DPPC diffusion with temperature from FRAP results, $T_m$ of the SLB can be determined.[6,13] On mica the $T_m$ value was 40.2±0.3 °C and is close to the Differential Scanning Calorimetry (DSC) measurement in free-floating DPPC Multilamellar Vesicles (MLVs) measured here, 39.73±0.02 °C (Figure S6) and in literature.[44,45] It should be noted that this is a DSC cooling scan value to match the FRAP data also collected during cooling, and that cooling rates cause an offset in $T_m$ values away from the true value (details in SI). This finding, that the $T_m$ of a lipid bilayer on mica is similar to free floating GUVs or MLVs, is also broadly replicated in other FRAP-with-temperature studies.[6,13,46]

The $T_m$ determined on glass was 38.6±0.2 °C, a small but statistically significant drop in $T_m$ of 1.6 °C from mica [Independent samples T-test, mica (M=40.2, SD=1.1), glass (M=38.6, SD=0.40), conditions; t(8.0)=4.64, p=0.002]. This implies a disordering of the lipid molecules within the bilayer on glass compared to mica. This is supported by literature observations showing a 1.4 °C drop in $T_m$ between the first and second bilayers in a double supported bilayer,[13] and a 2 °C drop in $T_m$ for bilayers on glass-like silica beads compared to MLVs.[44] In contrast, recent results have shown an increase in the $L_o$-$L_d$ miscibility $T_m$ when a GUV is ruptured onto a glass surface, implying an increase in order.[17] However these domains are pre-formed and show a demixing miscibility $T_m$ not a single lipid melting $T_m$ which could affect the interpretation of this result.

Further study is needed to link substrate effects with both single lipid $T_m$ and $T_m$ in phase separated mixtures, as well as examining the effects of bilayer deposition methods on these values.

Lipid diffusion on mica and glass supported SLBs is identical when we use the same lipid, the same dye, buffer, deposition technique, and equipment and analysis methods. The hypothesis being tested was that a short-range force originating in the chemical difference, higher on glass, was acting to slow lipid diffusion and thereby kinetically hinder the growth of nucleated domains. This is not the case.

## Substrate Roughness is Correlated with Domain Size

Roughness of all the substrates were measured using AFM to explore the possibility that surface roughness could be affecting domain formation (Table 1, Figure S8). The $R_a$ roughness of glass (0.15 nm) after piranha cleaning and UV ozone cleaning is over 4 times rougher than mica (0.03 nm) after cleavage (Figure 5A+B). These values match closely to previous AFM roughness measurements of mica,[15,37] and piranha cleaned glass.[14] We also studied other substrates used to support lipid bilayers, including glass with a differing composition (Glass 2), quartz, silicon and thermally oxidised silicon (Figure 2). Glass 2 and quartz also show order of magnitude higher roughness values 0.17 nm and 0.41 nm respectively (Figure S8) compared to mica, along with significantly smaller domains. $R_a$ however is a crude measure of roughness, devoid of in-plane information about the spatial frequency of surface undulations. It tends to report more strongly on out-of-plane deviations at longer wavelength. It might be that "roughness" at the scale where the bilayer can easily bend and conform does not have an influence, hence the need to quantify roughness at more relevant lengths.

Power density spectra quantifying the intensity of the out-of-plane deviations versus wavelength are shown for the substrate's 3D surface topography AFM images (Figure 5C and S8). Roughness is nearly two orders of magnitude larger for the glasses and quartz compared to mica across all length scales, reflecting the $R_a$. Silicon is somewhat rougher than mica across all wavelengths. Thermally oxidised silicon seems to be significantly rougher than silicon, but this is predominantly at several hundred nanometre wavelength and greater (Fig S8). At the smaller length scale <20 nm the roughness of thermally oxidised silicon is converging with that of silicon.

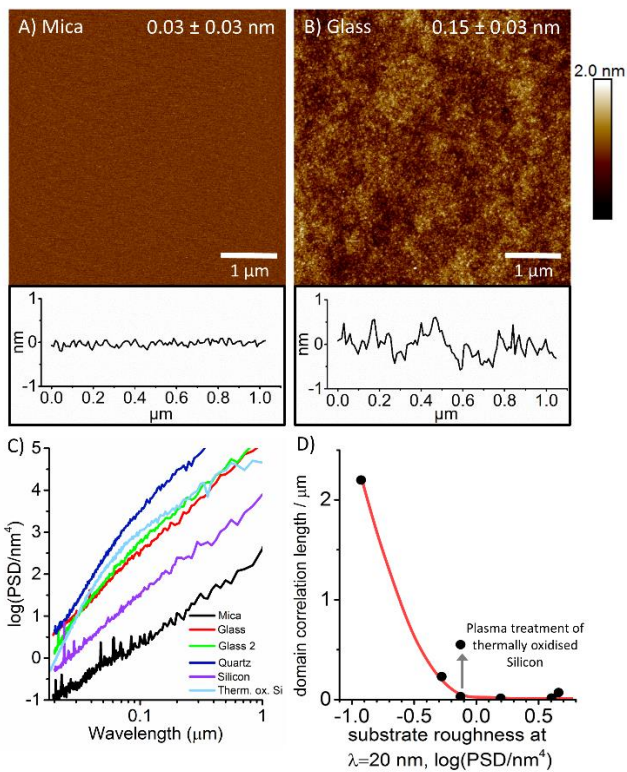

**Figure 5. AFM images of mica after cleavage (A) and glass after Pirnaha and UV ozone clean (B). The $R_a$ roughness measurements averaged over repeat images are included in top right of images. Individual AFM line scans are included below images. C is a Power Spectral Denisty Plot describing the roughness of different susbstrates at different length scales (see Figure S8 for images of the other clean substrates). The roughness trace for mica represents the noise floor of the AFM and laboratory used in this study, mica will be atomically smooth.**

**D shows the correlation length of domains with surface PSD roughness corresponding to a 20nm wavelength. Red line is a guide to the eye.**

If we measure the roughness of each substrate at the smallest length scale, and correlate with domain size, we find domain size rapidly diminishes as roughness increases (Figure 5D). Domains may grow to large sizes on mica, silicon and thermally oxidised silicon, but on the glasses and quartz they cannot grow. The behaviour on thermally oxidised silicon is interesting in that it is intermediate, behaving like glasses/quartz before plasma treatment, but like mica/silicon after plasma treatment. This seems to indicate that the domain formation is somewhat bimodal, with a roughness threshold above which domains cannot grow, and below which they are free to develop. Thermally oxidised silicon is around the level where effects of the surface chemistry are enough to switch behavior.

To confirm whether the difference in domain formation is largely due to roughness and not differences in surface chemistry interaction, a controlled degree of roughness was introduced to mica. Mica was treated using Hydrofluoric Acid (HF) to form 1.0 nm deep etch pits, defined by the crystallographic structure of mica (Figure 6A). The etched mica has an order of magnitude (at least) larger $R_a$ roughness than freshly cleaved mica, from $0.03\pm0.03$ nm (the noise floor of the AFM, the actual mica will be atomically smooth) to $0.26\pm0.03$ nm. Bilayers on etched mica produced much smaller domains than on flat mica (Figure 6B), with morphologies and correlation lengths (57 nm) much closer to the domains on glass. The roughness is higher than glass, but with a lower spatial density, but nonetheless proves that as the surface is roughened, large scale domain formation is dramatically hindered.

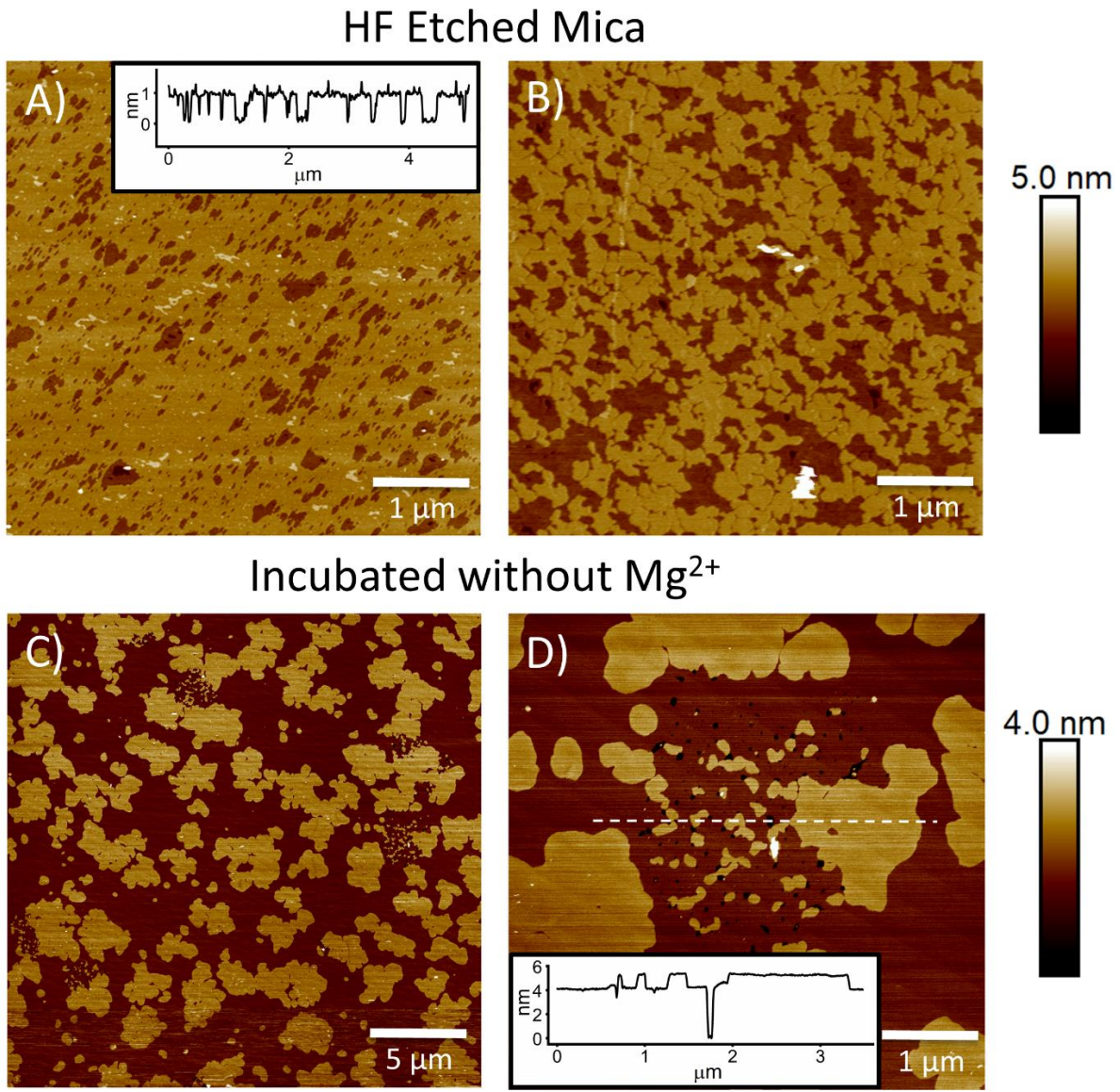

**Figure 6. AFM images of A) 1 nm pits in HF etched mica, and B) DPPC/DOPC(60:40) bilayer on the same HF etched mica. Note - A and B are not the same area. $R_a$ roughness of A is 0.26±0.03 nm. Correlation length of Domains in B is 57 nm. C) DPPC/DOPC(60:40) incubated as normal on mica but without the final $Mg^{2+}$ step used to aid vesicle fusion. Pin-hole defects are distributed across the image, with several areas of high defect density D) where domain growth has been halted by pinning to the defects.**

When the bilayer is prepared using vesicle fusion but omitting the 10 mM $Mg^{2+}$ (final diluted concentration), its formation is often incomplete, with pinhole defects across the image and areas of more serious disruption (Figure 6C). A DPPC:DOPC(60:40) bilayer prepared in this manner shows that domains can be pinned at the defects during the growth phase, restricting their ability to developing into larger domains. This points to the idea that the small domains can be explained by surface pinning.

# Discussion

The nanoscale domains formed on glass, quartz and thermally oxidised silicon substrates are orders of magnitude smaller than on mica, silicon and plasma-treated thermally oxidised silicon. Closer study of bilayer behaviour on exemplars of each type of behaviour, glass and mica, revealed insignificant or no differences in substrate hydrophilicity, molecular diffusion, and a slight reduction of $T_m$ on glass. Roughness at nanoscale wavelengths showed a strong correlation with domain size, with larger domains forming on smoother substrates. We have also demonstrated that mica roughened on the nanoscale hinders domain growth with a length scale similar to glass, indicating that it is the topography, not the chemistry, that is different between the substrates. What is the mechanism for the restriction of domain formation?

## Lipid Diffusion

Increased roughness has been shown to reduce diffusion up to 5-fold on etched glass and silicon surfaces ($R_q$=0.1-3 nm)[47] and silicon zerogels ($R_q$=0.71 nm)[48] due to curvature-induced areas of ordered bilayer, curvature-induced holes and hidden areas effects due to the 3D vertical component of diffusion. However these are at roughness values above the roughness of commonly used laboratory glass substrates (0.15-0.17 nm), on which there was no drop in diffusion compared to mica, shown by our results and by Benda et al.[12] Differences in molecular diffusion cannot explain the nanodomains.

## Lipid Order

AFM observations by the authors and others show that the top surface of bilayers are rougher on glass and silica compared to mica, the bilayer replicating the surface roughness to some degree.[15,37,48] Course-grained molecular dynamics simulations show that molecular scale corrugations 0.3 nm in height and width decrease the degree of periodic bilayer ordering.[49] These effects are far below the bending length scales of a membrane under the rules of continuum mechanics, where a bending rigidity of around 0.5-1 x $10^{-19}$ J[50] leads to metre scale persistence lengths,[51] or bending radii < 1 µm being energetically very costly.[52] Our experiments show that $T_m$ and hence lipid order was reduced slightly on glass (Figure 4B). The lateral dimensions of surface roughness are important. For 1-10 nm surface features the bilayer does not simply follow the surface curvature, and bilayers can span across pores that are less than twice the bilayer width (around 8-10nm).[48] This all suggests that if the roughness is a small percentage of the lipid bilayer height and the lateral separation between roughness peaks is small compared to bilayer dimensions, instead of curving to follow the corrugations of the surface, the roughness instead induces disorder in the bilayer. This would be a different mechanism to membrane bending, requiring deviations from the bilayer mid-plane. It is a distinct possibility that nanoscale surface roughness can disrupt the internal order within the bilayer.

Another more subtle possibility is that, as again shown in Figure 4B, the diffusion drops at a lower temperature on glass suggesting that the temperature at which the transition occurs is penalised by the surface, perhaps due to the roughness creating disorder as explained above. We are now at a temperature below which the domains would normally begin to nucleate, and it could be argued that the system is in a super-cooled state, whereupon the domains crash out quickly forming many small domains. Previously published data from our lab shows that fast cooling (20 °C/min and 50 °C/min) of a similar SM/DOPC system, which forces bilayers into super cooled states, results in domains crashing out that lose the characteristic fractal gel domain structure of slower cools, have a bimodal distribution, but most importantly are not nanoscale.[3] In this study, the rate of cooling through the $T_m$ is the same for both mica and glass, it is slow (0.25 °C/min) and there is only a 1.6 °C difference in $T_m$, therefore this mechanism cannot explain the nano-domains.

## Thermal History

Figure 4B shows that the drop in diffusion values for pure DPPC on glass and mica occur with a 1.6 °C offset, and if the temperature during cooling was held at an identical value but just below $T_m$ on both surfaces, then the diffusion rate at that temperature would be hugely different due to being at different points along the step function describing solidification of the bilayer. It could be argued that this very different value of D would lead directly to the different phase structure. However, here both bilayers on mica and glass go through the same thermal trajectory. The rate of cooling is the same, the rate of change of D is the same, and they are both falling from the same absolute value of D, so nucleation will be occurring when the diffusion values are the same, only 1.6 °C apart, and the diffusion rates around this point will also be relatively the same. Thus, the same structure should be expected. Further counter evidence is that $T_m$ on mica is at the slightly higher value, so its diffusion rate is the first to slow during cooling, so domain growth would be more restricted compared with the more fluid bilayer on glass *at that temperature.* This is the converse of the data. It should be noted that the thermal demixing transition for the DPPC/DOPC (60:40) is around 29-33 °C (see methods) and not at the $T_m$ of pure DPPC, pure DPPC diffusion was studied as domains obstruct the FRAP diffusion measurements.

## Roughness Induced Nucleation

There are many more domains on the rougher substrates, and another possibility is that this could be due to the roughness lowering the local activation energy for domain nucleation, providing more sites for nucleation. However, it has been established that solid phase domains should favour bilayers with low curvature,[5,18] avoiding areas of roughness induced curvature and hindering nucleation. This could be another cause of the slightly lower $T_m$ of pure DPPC on glass, it actually suppresses nucleation rather than encourages it.

## Roughness Increases Drag for Hydrodynamic Flow

In free-floating GUVs domain coalescence can continue until only one large domain remains if other coalescence hindering mechanisms are not operating, such as electrostatic repulsion of charged domains, or stress fields due to induced curvature around domains. Surface friction in SLBs however could slow the process of domain coalescence by increasing drag on the motion of lipid clusters or small domains. Surface friction will be proportional to the area of contact, and domain area scales with the square of radius, hence SLB domains, collective groups of many lipids, will rapidly experience a surface drag force as their size increase,[30,53] even though the individual lipid molecules can diffuse at the same rate on glass and mica. In other words, it is hydrodynamic flow of a bilayer that is vastly slowed. Simulations by Ngamsaad et al. have confirmed that restriction in hydrodynamic motion of domains results in restriction of domain growth to a limiting size.[54] These ideas are also supported by the decades old observation of Radler et al. who showed that the spreading velocity of a bilayer, i.e. large scale hydrodynamic flow, is decreased by over an order of magnitude on glass compared to mica.[55]

If a substrate is rough, then it could increase friction by two mechanisms. Firstly, AFM and Neutron Reflectometry show that bilayers conform to surface topography i.e. roughness and curvature.[47,48,56] More ordered domains colocalise at lower curvature regions in bilayers, due to the larger bending energy penalty for tightly packed gel domains to conform to a rough and curved surface compared to fluid domains.[5,18] A rigid planar domain would find it more difficult to move across any form of corrugated surface due to this bending penalty, or even on the low roughness surfaces here, due to the decrease in lipid order described above. Secondly, direct pinning of the more rigid domain could occur at high points on the surface below, as a ship beached on a single high-point on the sea-bed. Direct evidence that bilayer pinning can restrict domain growth is shown in Figure 6, where regions of many pin-hole defects lead to the solid phase domains becoming trapped at the defect edge, and domain growth is arrested in these regions. The proportion of solid phase is about the same, but domain coalescence has been hindered.

Correlation of pinned nanodomains with the underlying surface asperities would prove this mechanism, but it is difficult to obtain good quality images of the underlying surface through the bilayer in the same area with the same probes used to image the bilayer. For comparison, side by side images of the clean substrates and bilayers at the same scale are shown in Figure S9, but these are not of the exact same area.

## Surface Interactions

Both surface friction mechanisms described above will be mediated by the 0.3-1 nm interstitial water layer between the bilayer and substrate,[30,47] which if structured to some degree would mean pinning points do not necessarily need to be in absolute contact with the surface peak. The switch in behaviour between the thermally oxidised silicon before and after plasma points to this effect. This substrate is intermediate in roughness, someway between mica and glass, such that a change in surface chemistry will be enough to change the viscosity and thickness of the confined water, and have a measurable effect on friction mechanisms. Properties of this confined water should vary somewhat, between mica, glass, quartz and silicon due to chemistry. Different ionic strengths of buffers have been shown to affect bilayer-substrate interactions and diffusion likely through an interstitial water layer mediated electrostatic interaction,[46] so a similar interaction may be affecting domains. Harb et al. find DPPC diffusion on glass and mica to be the same when high ionic strength buffers are used, but it is faster on glass when ionic strength is low.[46] Whilst these are Langmuir-Blodgett deposited bilayers, it does point to the fact that interstitial water structure can strongly influence the supported bilayer dynamics. In this study all experiments are carried out in pure water.

Other factors include the polarizability of surface groups, for example quartz has a lower dielectric constant than silica, and Van der Waals forces, for example the Hamaker constant in water between two surfaces of mica is $2.0 \times 10^{-20}$ J and quartz is $0.60 \times 10^{-20}$ J [57]. In a medium with hydrocarbons interacting (as in this scenario) it is even possible for the Van der Waals forces to become repulsive due to retardation of the dispersive forces. In a recent paper Motegi et al. calculated the total interaction energies of mica and silica with the bilayer versus distance, finding them to be very similar, for silica at $-9.7 \times 10^{-6}$ J/m$^2$ (d = 1.9 nm) and for mica $-7.9 \times 10^{-6}$ J/m$^2$ (d=2.0 nm).[37] Silica is in fact slightly stronger. Another chemical difference is the acidity of surface silanol groups, which results in a change in charge density with pH, with a crossover point from positive to negative at the surface $pK_a$. In quartz the $pK_a$ of these groups depends upon orientation, with acidic out-of-plane silanol having $pK_a$ = 5.6, and less acidic in-plane groups having $pK_a$ = 8.6.[58] Hydrogen bonding might even occur between the silanol groups and the phosphate moiety of the lipid if the bilayer approaches closely enough, and this could be the initial interaction, pinning domains at asperities before physical contact is made. H-bond chains could alter the structure within the water layer and also create a longer-range bond between surface and bilayer. We have attempted to mitigate these effects by using roughened mica as a substrate and find nanodomains form with a correlation length of 57 nm, suggesting that roughness is the overriding factor.

If the strength of the surface interaction with the bilayer was strong enough, it would seem likely that the coupling to the lower (proximal) leaflet would overwhelm the inter-leaflet coupling, and the two leaflets would decouple. This has been observed previously, with differences in diffusion constants and thermal transitions being measured across leaflets under certain conditions.[4,59,60] However, in this study, only single thermal transitions were detected, and no decoupling was observed during fluorescence imaging of crystallising domains. It is possible that diffusion in the proximal leaflet could have been slowed, but our FRAP measurement were not designed to detect this, with the acquisition time being too short. Seeger et al. showed that decoupling occurred on mica, but not silicon oxide.[59] This result has recently been supported by Motegi et al. who used single molecule tracking of labelled lipid molecules to find a second slower diffusive component on mica in contrast to a single peak on silica.[37] Interestingly, the average combined leaflet diffusion constant was the same on both surfaces, reflecting our FRAP measurement. The interpretation in both above studies was that smooth mica allows close approach of the bilayer, hence Van der Waals forces become stronger, but on rough surfaces the bilayer is on average further away, resulting in behaviour closer to free GUVs. This is completely opposite to our data. We observe unhindered domain growth on mica, not glass, and the FRAP diffusion coefficient on both substrates is the same, and much lower than in a GUV.

An explanation could be that we are studying two different effects or mechanisms. A strong surface interaction results in bilayer decoupling and slower diffusion of the lower leaflet, but the restriction of domain growth is not governed by molecular diffusion, but by friction and pinning of the mobile clusters and proto-domains with the surface roughness, severely restricting their ability to coalesce. The hydrodynamic flow is restricted, much as the spreading velocity of a bilayer is 10 fold slower on glass than mica.[55] A more detailed consideration of surface effects is given in the SI.

## Implications

Lipid domains on glass that are beyond the resolution of traditional microscopy are clearly visible using AFM, which is probably why there is a disproportionately small number of publications reporting phase separating systems on glass, compared for example to the ubiquitous phase separation in giant vesicles,[5,17,26,35] and supported bilayers on mica.[2,3,24,25] Of the published work looking at phase separated systems on glass, we found only one paper by Honigmann et al. showing similar nanoscale domains on glass.[11] They observed distinct micron scale liquid ordered-liquid disordered ($L_o$-$L_d$) domains on mica using fluorescence, but the structure of the same lipids on glass was only resolvable using a combination of super-resolution STED and STED-FCS, visualising nanoscale domains averaging 90nm diameter. The study complements ours, by showing that the nanoscale domains formed on glass also occur for liquid-liquid phase separating systems as well as gel-liquid systems. Seu et al. and Burns et al. show domains on glass with DPPC/DOPC and DPPC/DOPC/Chol systems respectively that do not match with any domains we have observed in similar systems,[14,29] or with domains observed by Honigmann et al.[11] The reason for this discrepancy is unknown, but could be due to the chemistry of a particular glass used. There are many varieties of glass available from numerous manufacturers. These include soda lime and borosilicate glasses, which vary in silicate composition, and also bio-active glasses, which are designed to behave in a particular fashion.

Visible domains have been observed in GUVs deposited onto glass substrates as long as the domains were already present in the GUVs before deposition,[16,17,28] implying that interaction with glass kinetically traps a pre-existing phase structure at the moment of absorption. These domains do not reform on temperature cycling to sufficiently high temperatures,[16,17] but a 'speckle' pattern is observed.[16] Similarly, Langmuir-Blodgett/Langmuir Schaefer bilayers formed on glass show phase separation, provided there was phase separation in the initial monolayers before deposition.[16,26] Again, when the temperature is cycled the 'speckle' pattern is observed.[16] These results can be explained by the glass locking the phase structure in place, then restricting domain growth if heated above $T_m$ then cooled.

## Conclusion

To summarise, the most likely scenario is as follows. As the fluid bilayer cools, ordered domains begin to nucleate, whether crystallising or demixing, depending on the lipid compostion. Domain growth is diffusion limited, but also occurs by coalescence of smaller clusters and domains via hydrodynamic flow, the coordinated motion of a body of lipids. The different outcomes of phase separation in SLBs on different substrates are not caused by changes in molecular diffusion, but rather due to surface roughness at the sub-nanometre scale causing drag on hydrodynamic flow, in proportion to the domain contact area, i.e. to the square of the radius, so drag rapidly increases with domain size. The roughness could cause drag by two mechanisms, a) by inducing disorder in the supported bilayer, slowing motion of the more ordered phase which prefers a planar low-curvature bilayer, and b) direct pinning at singular asperities on the surface or step edges around pits. Evidence for the second pinning mechanism comes from the very irregular shape of the nanodomains, and from observation of restricted domain growth around bilayer defects. Domain structure appears to be bimodal, in that domains are either < 100 nm or micron scale, there is a tipping point after which domain coalescence is arrested. This binary distinction could be due to the bilayer being free if there is enough bulk water lubricating the interstitial gap between the bilayer and substrate, but if not then the domains (which are deeper than the fluid phase) are pinned at high points and suddenly hydrodynamic flow of the domains is arrested. Mediation by this interstitial confined water means that subtle differences in the exact chemical nature of the surface, and of any solutes in the water layer, will determine how the surface asperities interact with developing domains. Hence an intermediate rough surface, such as thermally oxidised silicon, can show nanodomains, but after treatment with an oxygen plasma the interstitial water layer thickens slightly, allowing micron scale domains to coalesce. It is an interplay between surface nanoscale roughness, and the influence of physico-chemical surface properties on the confined interstitial water layer.

Although glass is used ubiquitously for optical imaging purposes, it requires cleaning in harsh chemicals to render it clean enough and hydrophilic enough to support lipid bilayers, and after this treatment the surface is still molecular rough and heterogeneous. Mica in comparison, can be cleaved in a matter of seconds using tape/tweezers/scalpel, leaving a clean, hydrophilic and atomically flat surface. Mica is commonly used as an AFM substrate, but high-quality optical images of bilayers can also be obtained through mica if it is thin enough.

In vivo the cytoskeleton, a dense layer of actin filaments pinned to the membrane by protein interactions,[61] can act to restrict domain growth. Macroscopic optically resolvable phase separation is not observable in the plasma membrane of cultured mammalian cells, but is observable in cytoskeleton-free Giant Plasma Membrane Vesicle (GPMVs) induced from these same cells.[62,63] Pinning a minimal cytoskeleton to a phase separating SLB restricts the growth of micron scale domains and when the pinning sites are in both the $L_o$ and the $L_d$ phase, the nanoscale phase structure is strikingly similar to the phase separation we see on glass.[64] Experiments with GUVs,[65] as well as simulations,[66–68] also show how cytoskeleton-like pinning sites can restrict large scale domain formation. This all provides strong evidence that the presence of the cytoskeleton is a factor in restricting macroscopic phase separation. Substrates may actually be a more accurate model for in vivo membranes than free-floating membranes and could be tuned to a precise degree for more accurate and controllable models of the cell membrane.

## Associated Content

Figures showing domain size and correlation length methods (S1-S2), Gaussian blurring to mimic the diffraction limit (S3), nucleation of the DPPC + 0.5mol% TR DHPE domains shown in Figure 3 (S4), a wider set of DPPC/DOPC (60:40) AFM images (S5), DSC of DPPC MLVs (S6), FRAP-with-temperature raw images (S7), AFM images of all substrates (S8+S9) and AFM phase images of domains (S10). Table showing contact angle measurements of glass after successive cleaning. Extra information on correlation length analysis, Gaussian blurring to mimic the diffraction limit, effects of cooling rate on $T_m$ in DSC and FRAP, and substrate decoupling, as well as DSC experimental information.

The data that supports the findings of this study are openly available at https://doi.org/10.5518/721.

## Author Information

Corresponding Author

*s.d.a.connell@leeds.ac.uk

## Acknowledgment

The work was supported by EPSRC grant EP/J017566/1 'CAPITALS'. We acknowledge Dr Peng Bao for help with fluorescence and FRAP measurements, and Dr Mark Tarn for the HF acid etch.

**For Table of Contents only**

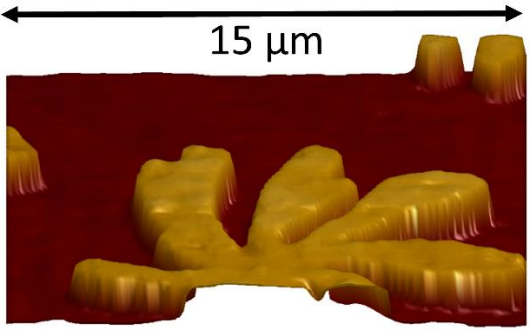 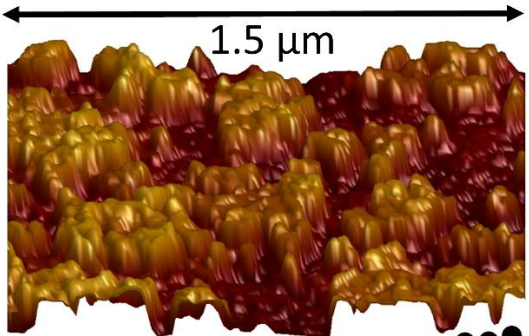

15 µm        1.5 µm

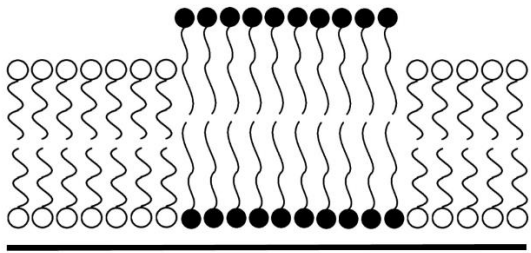 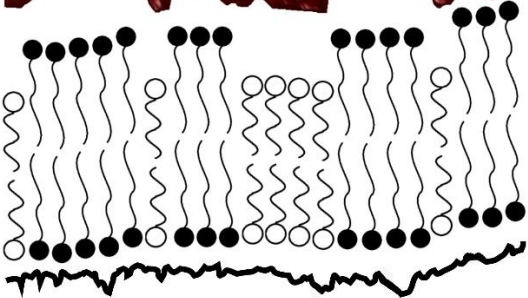

MICA SUBSTRATE        GLASS SUBSTRATE